\documentclass[structabstract]{aa}  
\usepackage{graphicx}
\usepackage{txfonts}
\newcommand{\lsp}{LS~I~+61$^{\circ}$303}

\newcommand{\lsi}{LS~I~+61$^{\circ}$303~}

\newcommand{\beq}{\begin{equation}}
\newcommand{\eneq}{\end{equation}}
\begin{document}
   \title{Feasibility Study of Lense-Thirring Precession in LS I +61$^{\circ}$303}

   \author{M. Massi 
          \and
          L. Zimmermann
          }

   \institute{Max Planck Institut f\"ur Radioastronomie, Auf dem H\"ugel 69, 53121
Bonn, Germany\\
              \email{mmassi@mpifr-bonn.mpg.de,  lzimmerm@mpifr-bonn.mpg.de}
             }
   \date{Received 2/12/2009; accepted 18/03/2010}

  \abstract
   {Very recent analysis of the radio spectral index and high energy observations have shown that the two-peak accretion/ejection microquasar model  applies for \lsp.}
   {The fast variations of the position angle observed with MERLIN and confirmed by consecutive VLBA images
must  therefore be explained in the context of the microquasar scenario.}
   {We calculate what could be the precessional period for the accretion disk in \lsi 
under tidal forces of the Be star ($P_{tidal-forces}$) or under the effect
of frame dragging produced by the rotation of the compact object ($P_{Lense-Thirring}$).}
   {$P_{tidal-forces}$ is more than one year. $P_{Lense-Thirring}$ depends
on the truncated radius of the accretion disk, $R_{tr}$. We determined $R_{tr}=300 r_g$ for observed QPO
at 2 Hz. This value is much above the few $r_g$, where the Bardeen-Petterson effect should align the midplane of the disk.
For this truncated radius of the accretion disk 
$P_{Lense-Thirring}$  for a slow rotator results in a few days.}
   {Lense-Thirring precession induced by  a slowly rotating compact object could be compatible
with the daily variations of the ejecta angle observed in \lsp.}


\titlerunning{Lense-Thirring Precession in \lsi}

   \maketitle
\section{Introduction}
\object{\lsi} is a X-ray binary formed by a compact object and a massive star with an optical 
spectrum typical for a rapidly rotating B0 V star (\cite{hutchings-crampton1981}).
The Be star  with a fast polar wind has an equatorial dense, low velocity wind with a
power law density distribution of the form $\rho_w (R)=\rho_0(R/R_*)^{-3}$, where $R_*$
is the radius of the primary Be star  (Waters et al. 1988; \cite{marti-paredes1995}).
The real nature of the compact object travelling in this stratified wind on an eccentric
orbit with e=$0.54-0.7$ is still unknown (\cite{aragona}, \cite{casares2005}). In fact, because of the uncertainty in the inclination of the orbit $i=30^{\degr}\pm20^{\degr}$,
  the  compact object could be either a  neutron star  or 
a black hole of 3-4 M$\odot$. Two radio periodicities are present (Gregory 2002), one of 26.5 d (phase $\Phi$), which corresponds to the orbital period, and a second one of 4.6 yr (phase $\Theta$) which is related to variations in the equatorial wind of the Be star (Zamanov \& Marti 2000). 
In the past two scenarios were presented for the system. One is that the
compact object is a young, still very  fast rotating, strongly magnetized pulsar, 
whose relativistic  wind collides with the Be star`s wind and prevents any accretion. 
In this scenario the prolonged strong wind interaction during the periastron passage  should
continuously
accelerate particles to relativistic velocities and 
a prolonged  optically thin outburst is expected,
as it occurs in the young pulsar PSR B1259-63 (Fig.~3 of   Connors et al. 2002).
PSR B1259-63, which  clearly is a pulsar, as seen by  its 
  pulses with a period of  48 ms  observed by Johnston et al. (1992), 
shows  a large optically thin radio outburst around the periastron passage.
The other scenario for \lsi is that the compact object is an accreting black hole or a low magnetic field 
 neutron star. From Bondi (1952) the wind accretion rate is  proportional to $\rho_w \over \rm v^3$, i.e.
directly proportional
to the wind density  and inversely proportional
 to the cubic of the relative speed 
between the compact object and the Be  wind.
 In an eccentric orbit this different relationship for density and velocity creates  
two  peaks in the accretion 
rate curve, one at periastron where the density is at its maximum
 and a second one when 
the drop in density is compensated by the decrease in velocity towards apastron.
Taylor et al. (1992) computed the accretion rate curve for different eccentricities and 
showed that two peaks begin to appear  for an eccentricity above 0.4.
Whereas the first peak is always toward periastron, the orbital
occurrence of the  second accretion peak 
 depends on variations of the wind of the Be star. 
Marti \& Paredes (1995) computed the accretion rate curve for different wind velocities,
associated with the variability of the Be star, and showed that for a stellar wind velocity of 
20 km/sec the two peaks become rather close to each other, whereas for a wind velocity 
of 5 km/sec they are 
at their maximum orbital offset of $\Delta \Phi=0.4$. This value, for an orbital period 
 P$_{orbital}$=26.5 d, corresponds to almost 11 days, i.e. the second peak may occur
almost at apastron.
Bosh-Ramon et al. (2006)  showed that around periastron ($\Phi$=0.23) only a small radio outburst is 
expected in coincidence with the first accretion peak. That occurs, because of the 
severe external inverse Compton (EIC) losses of the electrons, which upscatter UV stellar photons of the close Be star to high energies.
In other words, associated with the first accretion/ejection peak a high energy outburst due to
EIC is expected along with a small radio outburst.
The second displaced accretion/ejection peak, instead, should be observed in the radio band as
a large outburst.
In particular this large radio outburst, $S\propto \nu^\alpha$, should
follow the characteristics of microquasars:
optically thick emission, i.e.  $\alpha\geq 0$, followed by an optically thin outburst, i.e. $\alpha < 0$,
 (Fender et al. 2004).
The first type of emission, the optically thick radio emission, is related
in microquasars to a steady, low velocity jet centered on the orbit.
 The following  optically thin outburst
is  related to a transient jet, associated with shocks quite  
  displaced from the center (see Fig. 1).
With this transient jet  very high energy emission is expected as well, because of dominating self synchrotron Compton (SSC) losses during the growing 
phase of the shock (Marscher \& Gear 1985).

How does  \lsi fit in the two different scenarios? As a matter of fact, radio pulses have 
never been observed in \lsi, 
moreover  the large outburst in \lsi is clearly 
shifted towards  apastron contrary to the case of PSR B1259-63
(Table 1).
In addition, the recent analysis of the radio spectral index, $\alpha$, by Massi and  Kaufman Bernad\'o  (2009)
show the  clear sequence typical for  microquasars: Optically thick  emission (steady jet)
and optically thin emission (transient  jet). 
The quite impressive fact in \lsi is that during the maximum of the 4.6 yr periodicity,
i.e $\Theta=0.7-1.3$, this  sequence (optically thick  emission /
optically thin emission)
occurs twice along the orbit of 26.5 d. 
 In  Fig. 3 in Massi and  Kaufman Bernad\'o  (2009) one sees how  
even the small radio peak  at periastron, attenuated because of  severe EIC losses and nearly 
negligible in terms of flux density in comparison to the large $\Delta \Phi=0.3$ 
outburst,
presents the same  very clear variation in the spectral index, from $\alpha \geq 0$ (steady jet) 
to $\alpha <0$ (transient jet),
which proves the two-peak accretion model in the radio band.
   \begin{figure}
   \centering
\includegraphics[width=.35\textheight]{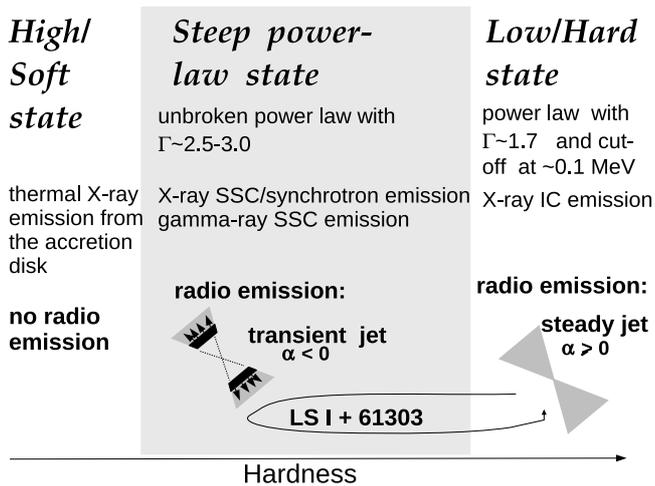}
      \caption{X-ray states/radio states vs. hardness. Along with  the usual X-ray characteristics 
(McClintock \& Remillard 2006) and radio characteristics 
(Fender et al. 2004)  of the spectral  states,  we add for the Steep power-law state also the X- and gamma-ray 
characteristics for the shock-in-jet model and for blazars 
(Marscher \& Gear 1985, Katarzy\'nski et al. 2005, Katarzy\'nski and Walczewska 2010).
The transient radio jet,
i.e. "plasmoids" displaced from the  center,
is related to shocks travelling in the slowly moving
preexisting steady jet established during  the previous low hard state. 
\lsi seems to evolve from  a very low low/hard state  to
a steep power-law state.  Then it develops again 
a  steady  jet centered on the system  (i.e. low/hard state) and never reaches
 the thermal high/soft state.}
         \label{FigS}
   \end{figure}

In terms of high energy emission as discussed above two peaks are predicted, each of them formed
by  different EIC and SSC contributions. At periastron one expects 
  dominant EIC contribution to the gamma-ray emission, due to the proximity of the Be star.
Towards apastron one expects the contrary: Strong gamma-ray emission due to SSC associated with the powerful transient jet
with additional EIC gamma-ray emission that depends on 
the distance from the Be star. 
Two emission peaks are confirmed by  observations at high energies.
One data set of EGRET 
shows evidence of the  periastron peak. 
A second data set  shows a hint of a peak at periastron and a second peak shifted towards apastron ($\Phi$=0.5) (Massi et al. 2005). 
The recent Fermi LAT observations  confirm the two peaks (Abdo et al. 2009). The Fermi light curve is characterized by a broad peak after periastron as well as a smaller peak just before apastron. TeV observations with VERITAS (Acciari et al. 2009) along with the strong emission at $\Phi$=0.5-0.9, first detected with MAGIC (Albert et al. 2006), also give marginal evidence for emission at $\Phi$=0.2-0.3.

\lsi fits then well the two-peak microquasar scenario. Its radio properties show  a
recurrent  switch  between  transient jet and steady jet, the two  radio states.
As discussed in Fender et al. (2004)
in microquasars the two radio states, steady and transient jet, 
are simultaneous  
with two X-ray states: The low/hard state 
and the steep power law state respectively (shown in Fig.1).
This seems to happen also in \lsi.
As analysed in Massi and Kaufman Bernad\'o (2009)
INTEGRAL (keV-MeV)  observations
by Chernyakova et al.  (2006)
with the typical photon index $\Gamma=1.4-1.8$ of the
low/hard state occur  
 at $\Theta$ and $\Phi$  where  optically
thick radio emission is observed, i.e. $\alpha \geq 0$.
INTEGRAL observations with   $\Gamma=3.6^{+1.6}_{-1.1}$,
are measured 
at  $\Theta$ and $\Phi$  where 
radio emission with $\alpha < 0$ is observed. 
Moreover, when TeV emission is detected in \lsi with Cherenkov telescopes 
the energy spectrum is always well fitted by a power law with a photon index $\Gamma \simeq 2.6$ 
(Albert et al. 2009; Acciari et al 2009) independently of changes in the flux level
as expected  
for the steep power law state, where the photon index is
a fundamental property of the state and not the luminosity
(see discussion in Massi and Kaufman Bernad{\'o} 2009).
 Of particular interest are the recent observations by  Anderhub et. al. (2009) probing correlated X-ray emission, attributed to synchrotron radiation,
 and VHE emission. We note that the corrected X-ray flux of $9\times 10^{-12}$erg cm$^{-2}$s${-1}$
and the VHE flux of  $11\times 10^{-12}$erg cm$^{-2}$s${-1}$ result in $F_{VHE}\propto F_X^{\eta}$
with $\eta$=0.99, in agreement with  the correlation observed in blazars, where $\eta$ is in the range $0.99-3$ (Katarzy\'nski and Walczewska 2010). The high energy spectra of TeV blazars are explained with a model where
relativistic electrons  accelerated in a shock
emit  synchrotron radiation up to  X-rays. A fraction of this emission
is upscattered to higher energies by the same population of the electrons (SSC) (Katarzy\'nski et al. 2005).
This agrees with the predictions of the shock-in-jet model invoked for the optically thin radio outburst of the transient jet described above
 and associated to the steep power-law state. 
Because of the correspondence between X-ray states and  radio states we suggest that the recurrent  switch  between  
transient jet and steady jet  observed in \lsi
corresponds to a  continuous switch
 between a low/hard state and a steep power-law state (Fig. 1).
The source is therefore pratically frozen in a permanent microquasar state, in fact it is always radio loud, never reaching, as other X-ray binaries
 the high/soft thermal state.

The peculiar aspect of \lsi, that we will deal with in this study, is its  short-term variability. 
Variability is the main characteristic of  blazars and micro-blazars,  where the shock is seen almost face-on
 and the Doppler factor is large.
\lsi was indicated since 2002 by Kaufman Bernad{\'o} et al. as a microblazar, because of   
  its one-sided jet radio morphology, typical of blazars, 
where the receding jet is Doopler de-boosted whereas the flux density of the approching jet is strongly amplified.

\begin{table*}
\label{table1}      
\centering          
\begin{tabular}{c c c c  }     
\hline\hline       
&PSR B1259-63 & Two-peak Microquasar & \lsi \\ 
\hline                    
radiopulses & yes & no & no\\  
$L_X$ [erg/sec] & 5 $\times 10^{34}$(a)& $10^{33.5}-10^{36}$(b)  & $(1-6) \times 10^{34}$(c)\\
Largest radio outburst& toward periastron (d) & toward apastron (e) & toward apastron (f) \\
Radio spectral index& $\alpha < 0$ (d)& $\alpha\ge 0$  then  $\alpha < 0$  
(g)&  $\alpha \geq 0$ then  $\alpha < 0$  twice per orbit (b)\\
Gamma-ray peak& toward periastron (h)&  EIC at periastron, dominant SSC toward apastron (e,i)
&
Two peaks: EGRET/VERITAS/FERMI (i)\\
\hline                  
\end{tabular}
\caption{
{\small Two-peak MQ vs Pulsar Model. a) Kaspi et al. 1995. b) Massi and Kaufman Bernad{\'o} 2009. c) Paredes et al. 1997. d) Connors et al. 2002.  e) Bosh-Ramon 2006. f) Fig. 2-c in Massi and Kaufman Bernad{\'o} 2009.
g) Fender et al. 2004. h) Aharonian et al. 2009.   i) see Sec. 1}.}       
\end{table*}

 \lsi shows strong morphological changes: or
the position angle of the one-sided jet  continuosly changes, or at some epochs the receding jet appears unattenuated 
(compare the  map of Taylor et al. 2000 with that of Dhawan et al. 2006 
 in Fig. 1 by Massi \& Kaufman Bernad{\'o} 2009).
Precession of the accretion disk (and therefore of the jet) 
causing the jet to point  closer to or farther away from the
line of sight would explain variable Doppler boosting and variations
in the position angle.
Precessing jets are well known in microquasars, the most spectacular one being that of SS433, the first microquasar (\cite{dubner}).
However, in \lsi the variations seem to be very rapid compared
to the 164-day precessional period of SS433.
The peculiarity of the  variations of \lsi is their short timescale.
MERLIN images revealed a surprising variation
of 60$\degr$ in position angle in only one day (Massi et al. 2004). 
  Dhawan et al. (2006)  measured in  VLBA images a rotation of the inner structure of
roughly 5$\degr-7\degr$ in 2.5 hrs, that is again almost 60${\degr}$/day.
Indeed, because of the difficulty to explain this puzzling variations in the context of microquasars 
the variations were  interpreted as due to a cometary tail of a pulsar.
Now, that the radio spectral index analysis and very high energy observations 
confirm  the two-peak microquasar scenario (see Table 1),  an investigation of the physical processes behind these fast variations 
is necessary. This is the aim of this work.

The most likely cause for precession of an accretion disk of a compact object is 
an assymetric supernova explosion of the progenitor. 
As a result the compact object could be tilted (\cite{fragile07}). 
In this case either the accretion disk is  coplanar
with the compact object, and therefore subject to the gravitational torque of the Be star or 
instead, the accretion disk is  coplanar with the orbit, but 
tilted in respect to the compact object, which induces Lense-Thirring precession if the
compact object  rotates. 
In this paper we  therefore examine these two possibilities  with the 
aim to quantify them and
to compare them with the observed short time scale of variations in \lsp.
   \begin{figure}
   \centering
\includegraphics[width=.4\textheight]{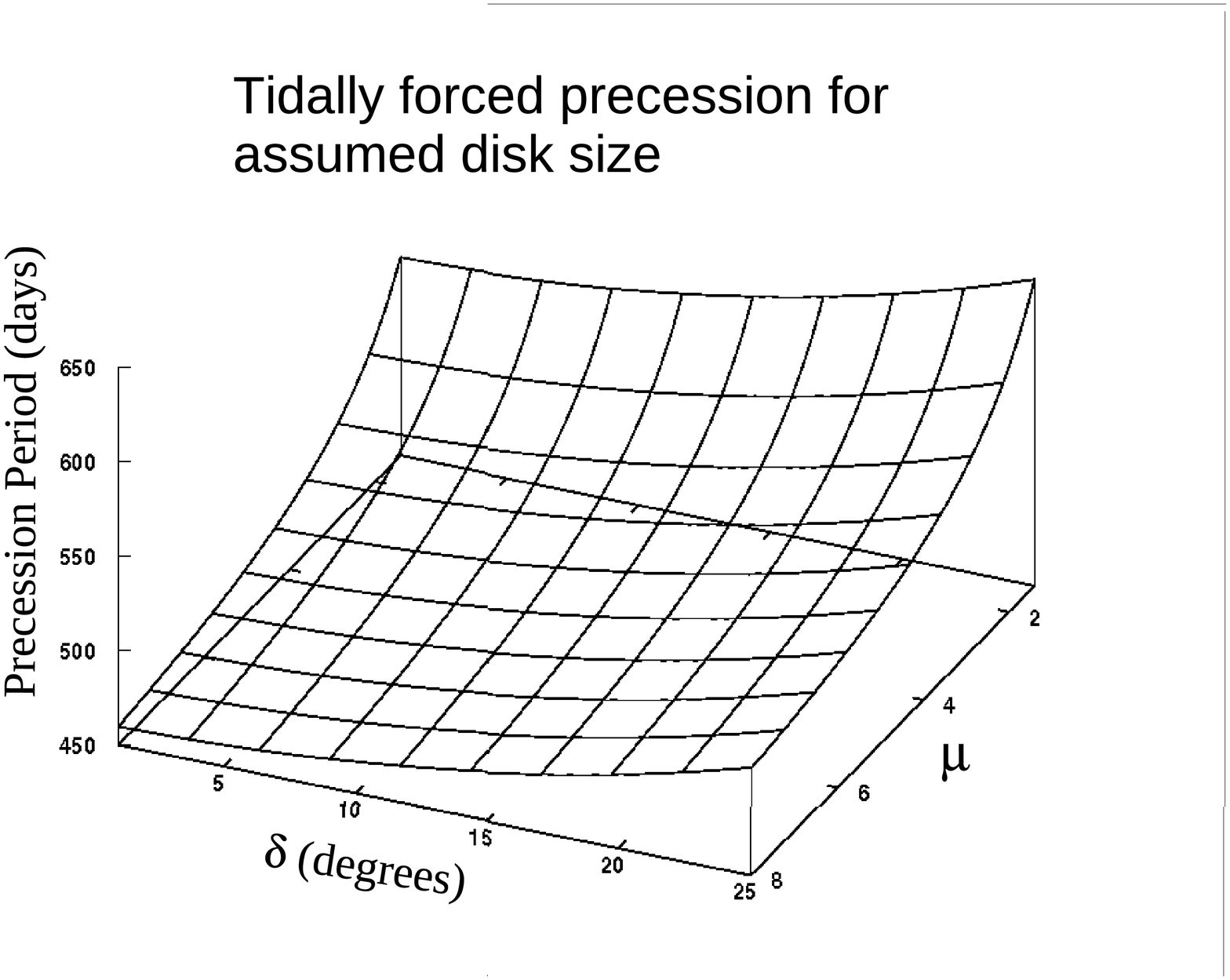}
      \caption{Precession due to tidal torque induced by the Be star as function of 
mass ratio, $\mu$, and  inclination angle , $\delta$,
of the orbital plane with respect to  the plane of the disk (Eq. \ref{larw2}).}
         \label{FigG}
   \end{figure}
   \begin{figure}
   \centering
\includegraphics[width=.4\textheight]{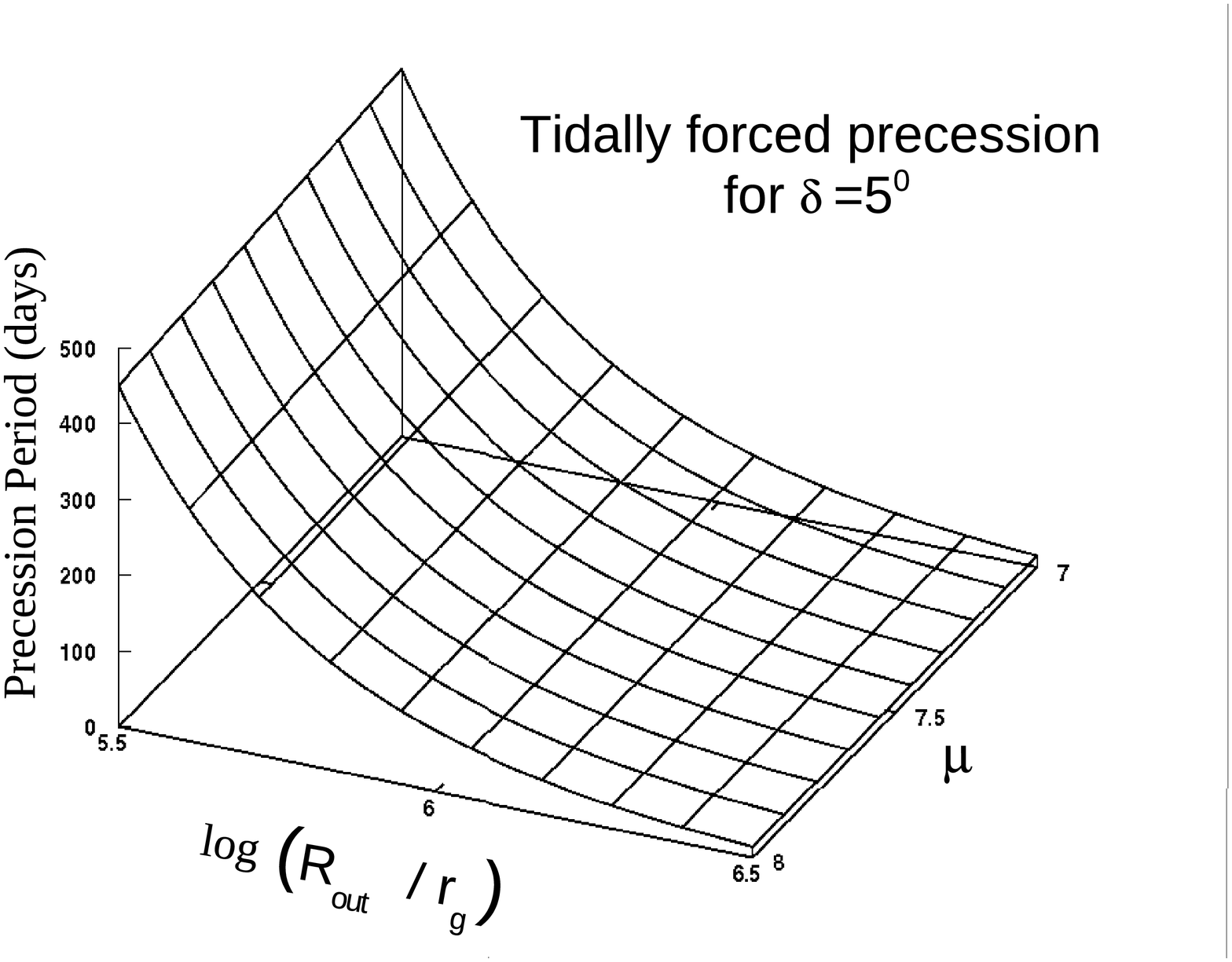}
      \caption{Precession due to tidal torque induced by the Be star  as function of
mass ratio, $\mu$, and disk size  
for inclination angle  $\delta=5^{\deg}$ (Eq. \ref{larw1}).}
         \label{FigG2}
   \end{figure}

\section{Tidal Forces in a Binary System}
If the accretion disk is tilted with respect to the binary orbital plane, then
its precession  can be tidally induced by the companion star. The expression for the
precession period depends, for a given orbital period ($P=26.496$~d) and semi-major axis ($a\simeq 10^{12}$~cm), 
on  the mass ratio $\mu={M_*\over M_{\rm{compact\,object}}}$, 
 the orbital inclination $\delta$ with  respect to the disk and the accretion disk size $R_{out}$,
(Larwood 1998, Eq. 4):
\beq
P_{\rm TF}={61.8 \over\cos\delta} {(1+\mu)^{1/2}\over\mu} ({a\over R_{out}})^{3/2}
\label{larw1}
\eneq
The size of the disk and the orbit can be eliminated, following Larwood (1998), by writing 
the accretion disc as a fraction $\beta$ of the Roche lobe. Then both $\beta$ and the Roche lobe can be written as a function of $\mu$, and Eq. \ref{larw1} then
becomes: 
\beq
P_{\rm TF}={61.8\over\cos\delta} {(1+\mu)^{1/2}\over\mu} [{0.6+\mu^{2/3}ln(1+\mu^{-1/3})\over 0.49 {1.4\over 1+[ln (1.8 \mu)]^{0.24}}}]^{3/2} 
\label{larw2}
\eneq
In \lsi the mass of the compact object is in the range
$2-3.5$ M$\odot$ and for the Be star one can assume the range $5-15$ M$\odot$, with a resulting
$\mu$ in the range $1.4-8$. 
In Fig.~2 we plot the precessional period, $P_{\rm TF}$, as a function of $\delta$ and $\mu$.
We see that even for the smallest values of $\delta$
and the highest values of $\mu$
we already get $P_{\rm TF}$= 460 d, which is more than one order of magnitude above the orbital period
of  \lsp.
This result of a period ratio ${P_{TF}\over P_{orbit}}\simeq 17$ agrees well with the expectations of the timescale for a tidally forced precession.
In fact, in Table 1 of Larwood (1998) one sees that the period ratio for known precessing X-ray binaries lies within the range
$8-22$.
However, such a long precessional period 
prevents tidal forces to be responsible for the fast variations of the position angle present in the MERLIN observations of Massi et al. (2004) and in the VLBA images of Dhawan et al. (2006). 

   \begin{figure}
\includegraphics[width=.37\textheight]{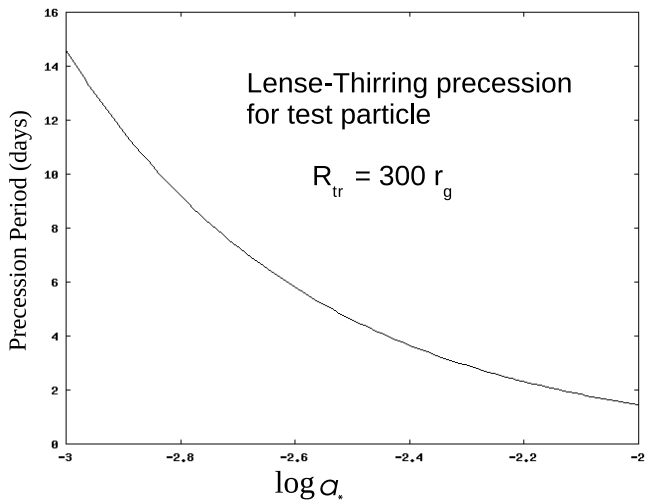}
      \caption{Lense-Thirring precession for a test particle  
 in function of the dimensionless specific angular
momentum, $a_*$, (Eq. \ref{LT1}). The orbit  of $R_{tr}= 300 r_g$ has been determined
by observed (Ray \& Hartman 2008) QPO at  2 Hz during the low/hard state.}
         \label{LT1fig}
   \end{figure}

   \begin{figure}
\includegraphics[width=.37\textheight]{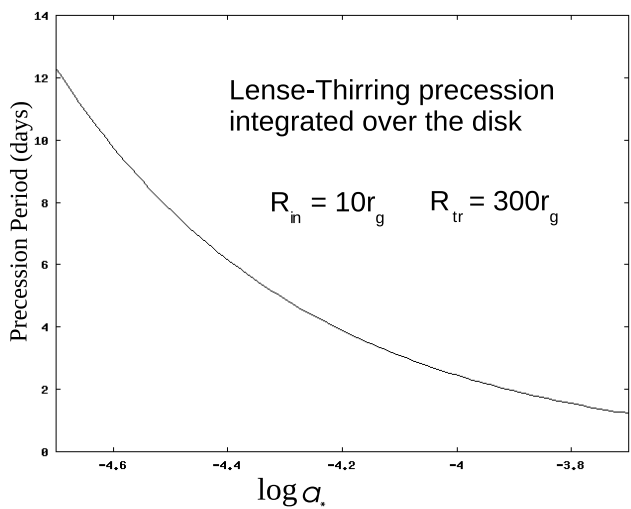}
      \caption{Lense-Thirring precession for a geometrically thick accretion flow  with inner and outer   radii
$r_i$ and $R_{tr}$ as in Eq. \ref{LT2} for  $\zeta=0.5$.}
         \label{LT2fig}
   \end{figure}

In our above calculations we  assumed a single planar disk.
If, instead, only a portion of the disk is warped out of the binary plane, as could be 
likely (Foulkes et al. 2006), a warp at smaller radii would
decrease $R_{out}$ in Eq. \ref{larw1} and further increase $P_{TF}$. The only way to get a smaller $P_{TF}$ is therefore to increase $R_{out}$.
In Fig. 3 we plot  Eq. \ref{larw1} as a function of $\mu$ and $R_{out}$ for $\delta=5\degr$. The first interesting result
is that the before determined value of $P_{TF}\simeq 450$~d by using Larwood's (1998) fit for $\beta$, which is derived from Paczy\`nski's (1977) values, corresponds to 
an already very large disk size, ${R_{out}\over r_g}=10^{5.5}$ (see Fig. \ref{FigG2}), 
where $r_g$ is the gravitational radius, $r_g=M_{\rm compact\,object} G/c^2$.
Indeed, Paczy\`nski (1977) argues that his values constitute upper limits to the disk size provided that pressure is sufficiently small in the disk.
One could therefore assume larger disks because of  higher values of pressure.
On the other hand, if we assume a larger disk size, we get the second interesting result that one needs to enlarge the disk size one order of magnitude,
reaching the  value of  ${R_{out}\over r_g}=10^{6.5}$, to finally reduce $P_{TF}$ to a few days.
Such a large $R_{out}$ value is unlikely. Already beyond about $10^4$ gravitational
radii, self-gravitation is larger than central gravitation, and the disk
becomes gravitationally unstable 
(Collin \&  Hur\'e 1999).

\section{The Lense-Thirring  Effect}
Whenever a spinning compact object has a misaligned accretion disk, the
Lense-Thirring precession effect will arise with an angular frequency,
$\Omega_{\rm LT}$, given by (Eq. 29 of Caproni et al. 2006; Wilkins 1972)
:
\beq
\Omega_{\rm LT}={2 \pi \over P_{\rm LT}}={2 G J\over c^2 r^3}
\eneq
\noindent 
where $J$ is the angular momentum of
the compact object  and $r$  the radial distance from the rotating compact object to the accretion disk.
The angular momentum, $J$, can be expressed in terms of the dimensionless spin parameter,
$a_*$, such that
${J={a_*GM_{\rm compact\,object}^2\over c}}$ (Fragile et al. 2001),
and the precessional period (in days) becomes
\beq
P_{\rm LT}= {1.8 \times 10^{-10}\over a_*} {M_{\rm compact\,object}\over M_{\odot}} \left({r\over r_g}\right)^3
\label{LT1}
\eneq

In general, the Bardeen-Petterson effect predicts an alignment of the inner accretion disk 
with the symmetry plane of the compact object.
Nelson  \& Papaloizou  (2000) showed that the alignment by the Bardeen-Petterson effect 
extends only to a few gravitational radii (15-30 $r_g$) from the compact object.
The rest of the disk remains tilted and therefore under the influence of the Lense-Thirring precession effect (Nelson  \& Papaloizou  2000).
The models of the low/hard state give  a geometrically thin, optically thick accretion disk truncated 
at   $R_{tr}\simeq$100 $r_g$ (\cite{mcclintock-remillard2006})
 and therefore well above the  limit of a few $r_g$.
In addition, for \lsi  the truncated radius  $R_{tr}$ is expected to be even larger than $r\simeq$100 $r_g$ because its state
with $L_X\simeq (1-6) \times 10^{34}$ (Table 1) corresponds to a very low low/hard state 
just  above  the upper limit of a quiescent state  
($L_{\rm X} (quiescent~state): 10^{30.5} - 10^{33.5}$ erg/s \cite{mcclintock-remillard2006}).

Following Stella \& Vietri (1998) we can argue
that matter inhomogeneities present at an inner
disk boundary   cause quasi-periodic oscillations (QPO)  at the
Keplerian frequency $\nu_K$. We can estimate the value for  $R_{tr}$ in \lsi from observed QPO.
During their monitoring with  RXTE  (2-10 keV) Ray \& Hartman (2008) 
observed a period of strong variability with a spectrum best fit
by a powerlaw of photon index about 1.5 (i.e. low/hard state);
a power spectral analysis revealed QPO at 2 Hz. The relativistic Keplerian frequency (Eq. 2 in Caproni et al. 2006) is:
\begin{equation}
\nu_K= {1 \over 2 \pi} {c \over r_g} \big[ ({R_{tr}\over r_g})^{3/2}+a_* \big ]^{-1},
\end{equation}
By solving for ${R_{tr}\over r_g}$ one determines:
\begin{equation}
{R_{tr}\over r_g}=\big[ {32\times 10^3 \over {{M_{\rm compact\,object}\over M_{\odot}} \nu_K}}-a_* \big ]^{2/3}
\label{kp}
\end{equation}
Equation \ref{kp} for  $M_{\rm compact\,object} = 3 M_{\odot}$ and $\nu_K=2$ Hz clearly
reduces to the Newtonian value for any value of $a_*$, and therefore to:
${R_{tr}\over r_g}= 300$.

Above, to determine the truncated radius of the accretion disk 
we set the QPO frequency of 2 Hz equal to the Keplerian frequency. 
Below, we analyse if the resulting ${R_{tr}\over r_g}= 300$ could correspond
to a precessional period of a few days.
This is therefore, different from Ingram et al. (2009) and Ingram \& Done (2009),
who associate the low frequency QPO directly to Lense-Thirring precession. 
As a matter of fact 
a truncated radius, $R_{tr}$, a factor 3 above the value of 
100 $r_g$ determined for X-ray binaries
in low/hard states, is very consistent with the X-ray luminosity value of \lsp, which indicates a very low low/hard state.
Moreover, the escape velocity $v/c=  {\sqrt {2 r_g \over r}}$ for $r=300 r_g$ 
results in $v/c=0.08$. 
Following Meier (2005) the terminal velocity of the  steady  jet, during the low/hard state,
 is approximately equal to
the escape speed at the footpoint of the magnetic field where the jet is launched, that is at the 
inner truncated radius of the accretion disk
(Meier 2005).
The value of $v/c=0.08$ should therefore be comparable with the velocity of the steady jet. 
Indeed, when \lsi was essentially quiescent Peracaula et al. (1998) measured an expansion velocity of  
$0.06\pm 0.01$.
 Therefore, on the basis of these consistencies we will assume in the following that
the determined value of  300 $r_g$  is a good estimate, in terms of order of magnitude, 
of the truncated radius during the rather low low/hard state of \lsp,
even if the Ray and Hartman observations of the QPO during a low/hard state of \lsi
are  at another epoch than the MERLIN and VLBA observations of the  precessing jet. 

In Fig.~\ref{LT1fig} we show the precessional period (Eq. \ref{LT1}) due to the Lense-Thirring effect as a function of
the dimensionless spin parameter $a_*$ for the determined radius $R_{tr}=300 r_g$. 
As a result we deduce a  period of a few days for the Lense-Thirring precession 
of a slow rotator ($0.001 < a_*<  0.01$). Therefore,   Lense-Thirring precession
could explain the  MERLIN and VLBA observations. 
 However, Eq.  \ref{LT1} assumes a single-particle orbit at the truncation radius
 ${R_{tr}\over r_g}= 300$, 
how does that change the result when we consider Lense-Thirring precession 
of the whole geometrically thick, advection-dominated  flow (ADAF) interior up until the truncated disc?
An expression of Lense-Thirring precession for disks was first given 
in Liu \& Melia (2002) and subsequently
reproduced in slightly different forms in Fragile et al. (2007) and Ingram et al.
(2009). Equation 2 in Ingram et al. (2009) solved for the precessional 
period (days) gives:
\beq
P_{\rm LT}= {1.2 \times 10^{-15} (1+2 \zeta) ({R_{tr} \over r_g})^{(2.5 -\zeta)} ({r_i \over r_g})^{(0.5 +\zeta)}(1-{r_i\over R_{tr}}^{(2.5-\zeta)})r_g \over a_* (5-2 \zeta)(1-{r_i\over R_{tr}}^{(0.5+\zeta)})}
\label{LT2}
\eneq

The surface density profile through the disk depends on $\zeta$ that 
for ADAF disk is $\zeta=0.5$  (Ingram et al.  2009).  
As $r_i$ we have to  assume the most inner radius, where  the flow remains  misaligned
and therefore subject to the Lense-Thirring precession, that is $r_i$
must be  clearly outside the Bardeen-Petterson effect.
As said above 
 Nelson \& Papaloizou (2000) determined the limit of 
(15-30) $r_g$. 
However, Nelson \& Papaloizou (2000) 
used in their  study a tilted geometrically thin accretion disk;
Fragile \& Anninos (2005) and Fragile 
et al. (2007) have shown that in case of  a geometrically thick disk the  Bardeen-Petterson limit is even
closer to the compact object and equal to   $r_i \sim 10 r_g$. 
In Fig. \ref{LT2fig}  we plot Eq. \ref{LT2} for $r_i=10$, $R_{tr}=300$ and $\zeta=0.5$.
The result of  Fig. \ref{LT2fig} is:  if the whole ADAF disk is precessing  $P_{LT}$ decreases.
In order to  mantain a precessional period of some days,
as calculated  for the single orbit above, it is enough to assume a slower rotating compact object, i.e. with lower  $a_*$. 
\section{Conclusions}
Two consecutive MERLIN observations of \lsi showed a rotation
of the position angle of the  radio structure of $\simeq 60\degr$
in only 24 hours (\cite{massi2004}). Several consecutive VLBA images by Dhawan et al (2006),
 three days apart, have confirmed the fast variations.
In this paper we analyse precession due to the tidal torque induced by the Be star and
to the Lense-Thirring precession induced by the  tilted rotating compact object.

It is unlikely that the observed  {\it days} time scale could be created by tidal 
precession. In this paper we show that this  mechanism would produce   too 
large a precessional period of $P_{\rm TF}\geq 460$d.
To lower the  precessional period to a few days one should increase the disk size above the
limit of any stable disk. 

On the contrary, for  Lense-Thirring precession for a single-particle orbit we determine
that a slow rotator, with $0.001 < a_* < 0.01$,
induces  a  $P_{\rm LT}$ of a few days for a truncated  radius  $R_{tr}=300\,r_g$. 
The large truncated radius,   $R_{tr}$, that we derived   from  
 QPO  observed with RXTE (Ray \& Hartman 2008),
is consistent with the  low/hard state of \lsi during those observations
(spectrum best fit by a powerlaw of photon index about 1.5; Ray \& Hartman 2008).
Moreover, the escape velocity, that following Meier (2005) is equal to the velocity of    
the steady jet,  for $R_{tr}=300 r_g$ results in  $v/c=0.08$ and  is therefore consistent with the expansion velocity of
$0.06\pm 0.01$ observed  by  Peracaula et al. (1998) in an epoch 
 when \lsi was essentially quiescent.

Taking into account the precession of the whole hot, inner thick ADAF disk (between  
the Bardeen-Petterson limiting radius of  10$r_g$ 
and the truncated radius of 300$r_g$) we obtain an even smaller  $P_{\rm LT}$.
In order to  mantain a precessional period of some days,
as calculated  for the single orbit, it is enough to assume a slower rotating compact object, i.e. with lower  $a_*$. 
We conclude that the Lense-Thirring mechanism can be applied to \lsi and so  explain
 the observed fast variations.

\begin{acknowledgements}
We are grateful to the anonymous referee for the constructive comments that improved the paper
and to Marina Kaufman Bernad{\'o}  and  Johannes Schmid-Burgk for comments and  suggestions.
The work of L. Zimmermann is partly supported by the German Excellence Initiative via the
Bonn Cologne Graduate School.
\end{acknowledgements}

\end{document}